\documentclass{PoS}

\def\met{$E_T^{miss}$\,}
\def\dpm{$\Delta{\hat{\phi}}_{min}$\,}

\title{Search for Natural SUSY with inclusive search strategies at the LHC using the CMS detector}

\ShortTitle{Inclusive searches for Natural SUSY with CMS}

\author{\speaker{Sezen Sekmen for the CMS Collaboration}\\%\thanks{A footnote may follow.}\\
        Physics Department, CERN, CH-1211 Geneva 23, Switzerland\\
        E-mail: \email{ssekmen@cern.ch}}

\abstract{Natural SUSY suggests the existence of light stop quarks, accessible at the LHC, which are the focus of a dedicated CMS search program.  I present two inclusive CMS searches that look for TeV scale colored sparticles in final states with jets, $b$-tagged jets and missing transverse energy performed using up to 19.4\,fb-1 of 8\,TeV LHC proton-proton data.  No deviation from the Standard Model was observed in these searches, and the implications for this was shown for several simplified model scenarios and phenomenological MSSM.}

\FullConference{The European Physical Society Conference on High Energy Physics -EPS-HEP2013\\
		18-24 July 2013\\
		Stockholm, Sweden}

\begin{document}

\section{Introduction}

Supersymmetry (SUSY) is a generic framework which can be realized in a huge diversity of final states.  In this vastness, we seek guidance of phenomenologically well-motivated SUSY scenarios when designing the SUSY searches at the LHC.  A scenario we thoroughly investigate nowadays is SUSY with natural EWSB, where the fine tuning in the loop corrections to Higgs mass are minimal.  Naturalness requires particles that couple to Higgs to be light, in order to avoid fine funing.  This constrains Higgsinos,  which contribute at leading order, to be ${\cal O}(100 {\rm GeV})$; stops, which contribute via one-loop corrections (and sbottoms which are related to stops) to be below a TeV, and gluinos, which contribute via two-loop corrections to be below a few TeV.  There are no constraints on the rest of the sparticle spectrum.

The Compact Muon Solenoid (CMS) Experiment at the LHC~\cite{Chatrchyan:2008aa} looks for Natural SUSY with a rich variety of searches, some of which look for stops in specific final states, and others look for colored sparticles in inclusive final states.  In the following, I briefly describe two inclusive searches that probe pair production of colored sparticles in final states with jets $b$-tagged jets and missing transverse energy (\met) performed using 8\,TeV LHC proton-proton collisions.

\section{jets, $b$-jets, $E_T^{miss}$ with variables $H_T$ and $\alpha_T$}

\begin{figure}[htbp]
\begin{center}
\includegraphics[width=0.50\linewidth]{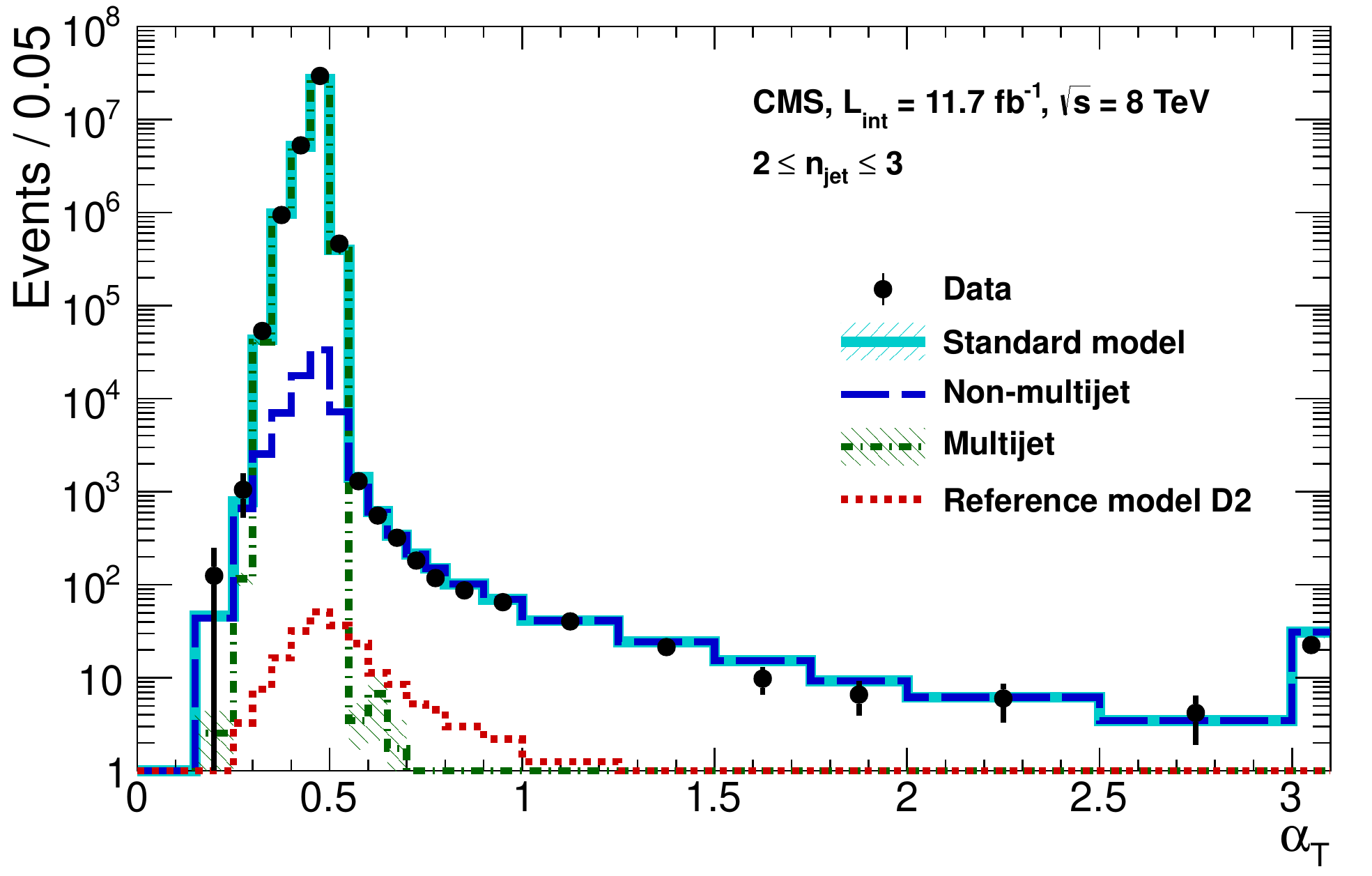}
\caption{The $\alpha_T$ distribution for events containing either two or three reconstructed jets. The reference signal model (labelled D2) is the direct pair-production of bottom squarks, which results in a final state containing two $b$-quarks and two LSPs. }
\label{fig:alphaT}
\end{center}
\end{figure}

The first search~\cite{Chatrchyan:2013lya} was performed using 11.7\,fb$^{-1}$ of data, and utilizes the kinematic variable $\alpha_T$, which provides discrimination between events with genuine \met from invisible particles, and misreconstructed \met from detector inefficiencies.  For a dijet system, $\alpha_T = E_T^{j2} / M_T$ where $E_T^{j2}$ is the transverse energy of the less energetic jet, and $M_T$ is the transverse mass of the dijet system.  When there are more than two jets in an event, $\alpha_T$ can be generalized by constructing a pseudo-dijet system, which was done here by choosing the combination that minimizes $\Delta H_T$ between the two pseudojets (where $H_T$ is the hadronic transverse energy).  For a perfectly measured dijet event having back-to-back jets with $E_T^{j1} = E_T^{j2}$ and no \met, $\alpha_T = 0.5$.   If an inbalance is caused in the back-to-back system due to a jet mismeasurement, $\alpha_T$ is less than 0.5.  In the case the dijet system is not back-to-back and recoils against genuine \met , $\alpha_T$ should be greater than 0.5.   Figure~\ref{fig:alphaT} shows the behavior of $\alpha_T$ for data, for multijet backgrounds with misreconstructed \met , for multijet backgrounds with genuine \met and for a reference signal model, and confirms the behavior of $\alpha_T$.  

Data used in this analysis were collected using triggers that require both $H_T$ and $\alpha_T$ to be above certain thresholds.  The baseline selection requires events to have at least two jets that have $p_T > 50$\,GeV and $|\eta| < 3$ and pass the jet identification criteria.  First two jets are required to have $p_T > 100$\,GeV.  Events are vetoed if non-identified jets with $p_T > 50$\,GeV and $|\eta| < 3$ are present.  To reduce SM backgrounds with genuine \met , events having isolated electrons, muons and taus are vetoed.  Events with isolated photons are also vetoed to ensure all-jet final states.  Finally $\alpha_T > 0.55$ is required to select signal-like events.  The analysis looks for a signal excess in a binned 3D space defined by jet multiplicity, $b$-jet multiplicity and $H_T$.  Jet multiplicity is binned as $n_j = 2-3, > 3$ in order to discriminate among $\tilde{q}\tilde{q}$, $\tilde{q}\tilde{g}$ and $\tilde{g}\tilde{g}$ final states whereas $b$-jet multiplicity is binned as $n_b = 0, 1, 2, 3, >3$ to provide sensitivity to 3rd generation.  $H_T$ is divided in 8 bins to probe different models spanning a large sparticle mass splitting range.  

The selected sample is still contaminated by SM backgrounds that need to be estimated.  Non-multijet backgrounds $tt+{\rm jets}$, $W(\rightarrow l\nu,\tau\nu)+{\rm jets}$ and $Z(\rightarrow \nu\nu)+{\rm jets}$ are estimated using control samples $\mu+{\rm jets}$, $\mu\mu+{\rm jets}$ and $\gamma + {\rm jets}$ which have negligible multijet background and signal contamination.  Expected background in the signal sample is obtained by multiplying the observed data in the control sample with a bin-dependent transfer function derived using Monte Carlo (MC) simulation.  Multijet backgrounds are estimated by exploiting the $H_T$-dependence of the ratio of number of events with $\alpha_T > 0.55$ over number of events with $\alpha_T \le 0.55$ as a function of $H_T$, which can be expressed as $Ae^{kH_T}$.  This analytical form was validated in a multijet-enriched data control region.

\begin{figure}[htbp]
\begin{center}
\includegraphics[width=0.35\linewidth]{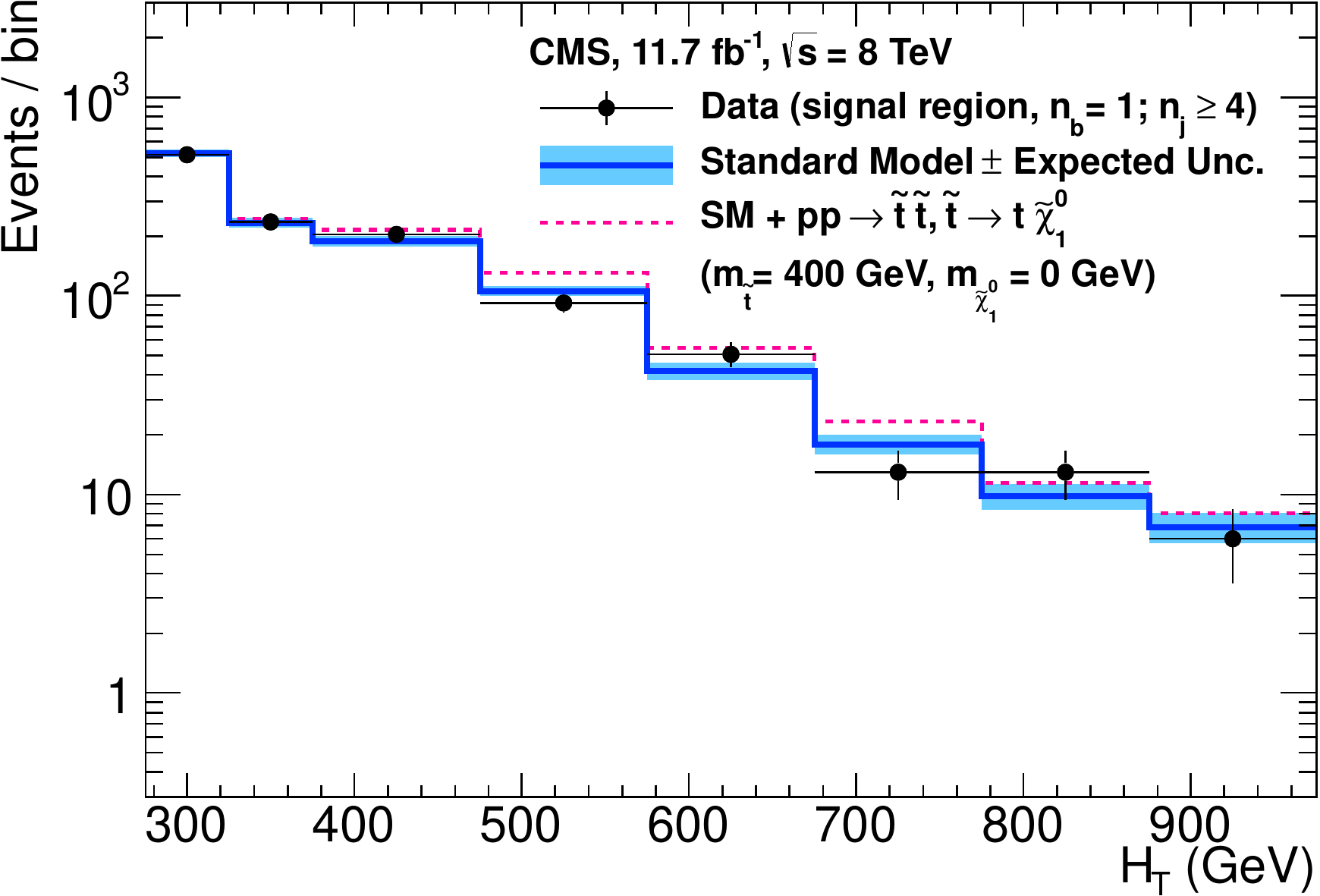} \quad
\includegraphics[width=0.35\linewidth]{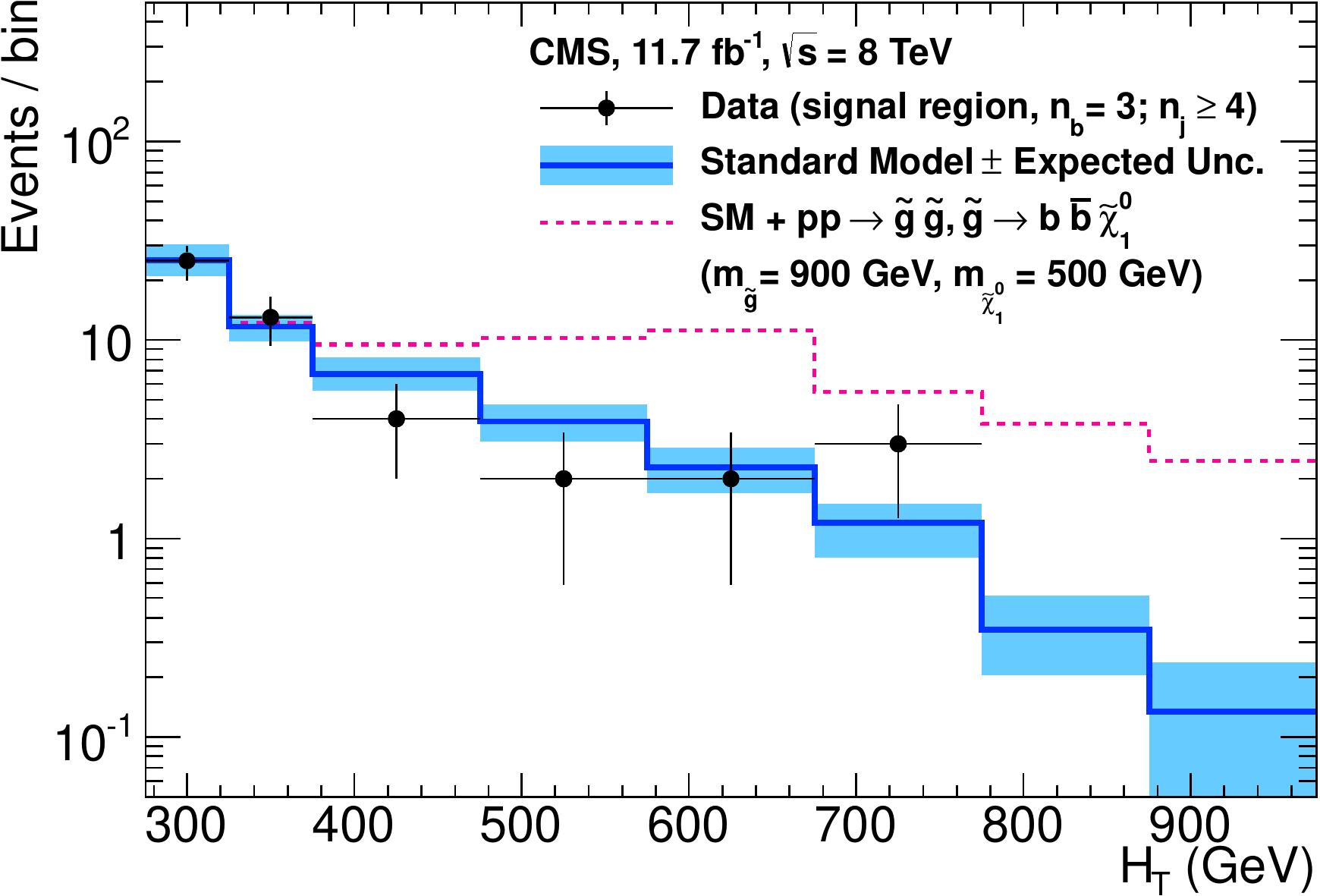} 
\caption{Comparison of the observed yields and SM expectations given by the simultaneous fit in bins of $H_T$ for the signal samples when requiring $n_j \ge 4$ and $n_b = 1$ (left) and $n_j \ge 4$ and $n_b = 3$ (right).  The example signal model shown  comprise pair-produced 400 GeV stops decaying to top plus a 0 GeV neutralino LSP. }
\label{fig:RA1fit1bge4j}
\end{center}
\end{figure}

For a given category satisfying jet and $b$-jet multiplicities, a likelihood model of the observations in multiple data samples was used in order to obtain a consistent prediction of SM backgrounds and for the presence of a variety of the signal models.  The likelihood analysis showed that data was consistent with the SM.  Figure~\ref{fig:RA1fit1bge4j} depicts this consistency for the $n_j \ge 4$ and $n_b = 1$ selection for the signal and control samples.  

\begin{figure}[htbp]
\begin{center}
\includegraphics[width=0.35\linewidth]{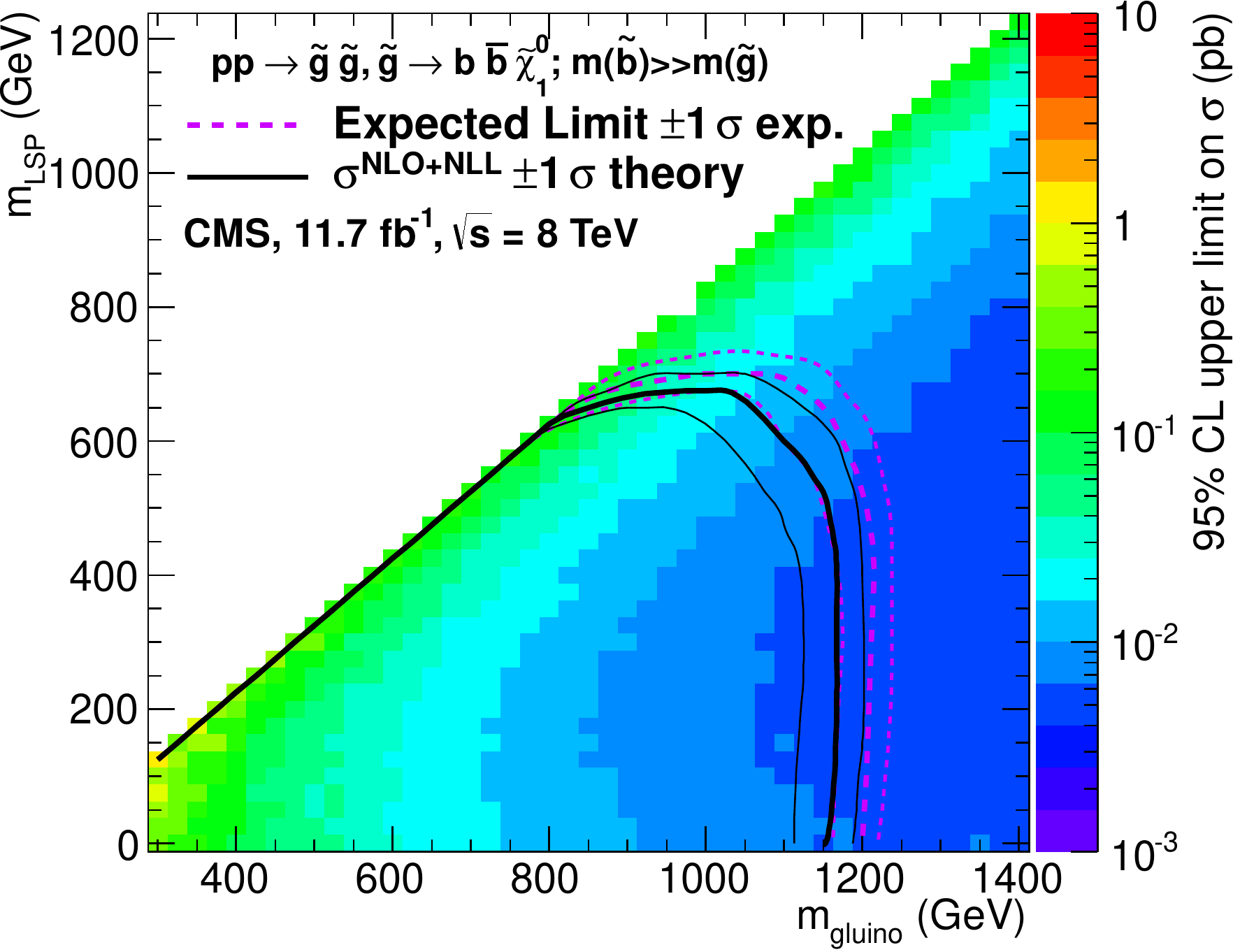} \quad
\includegraphics[width=0.35\linewidth]{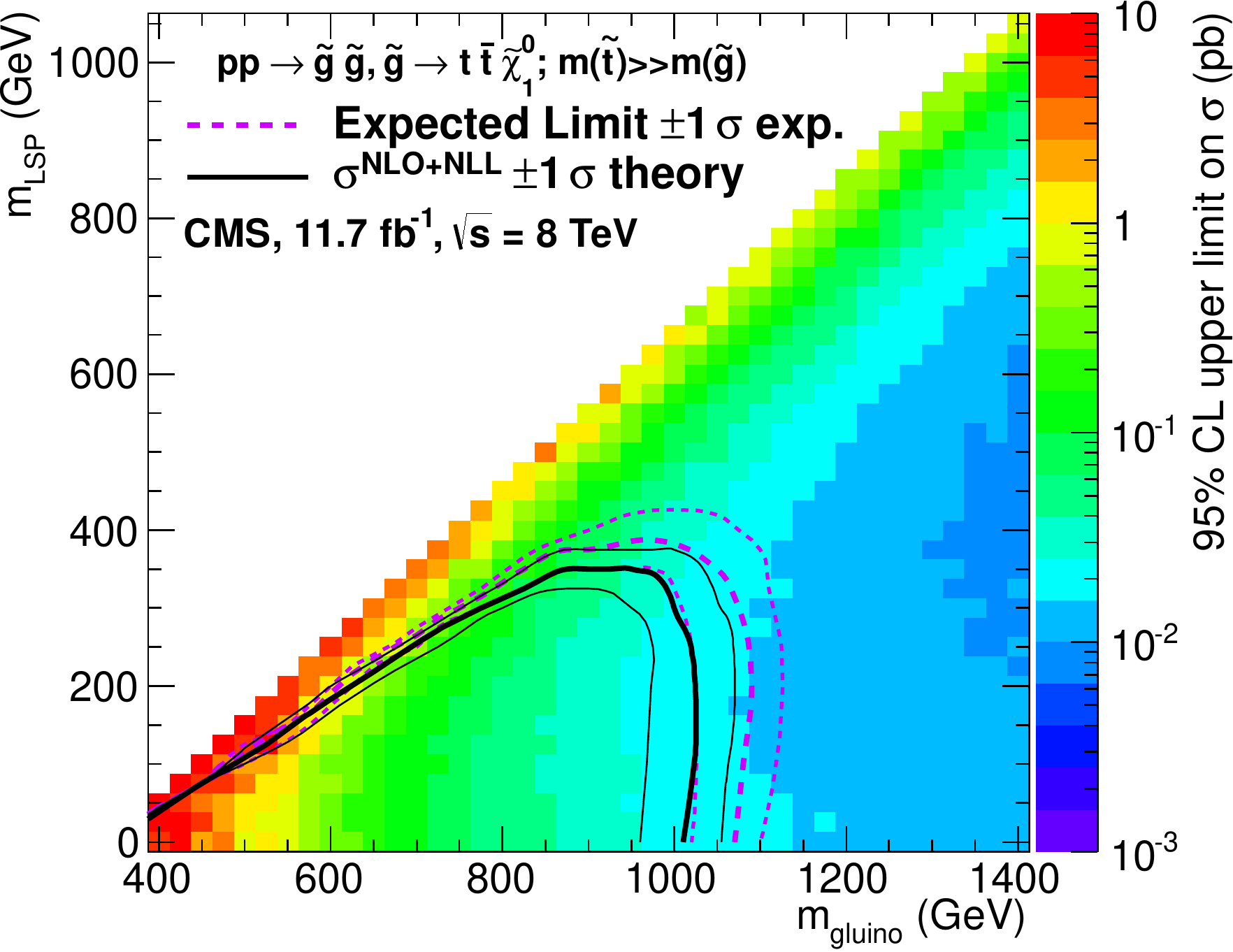} \\ \vspace{0.2cm}
\includegraphics[width=0.35\linewidth]{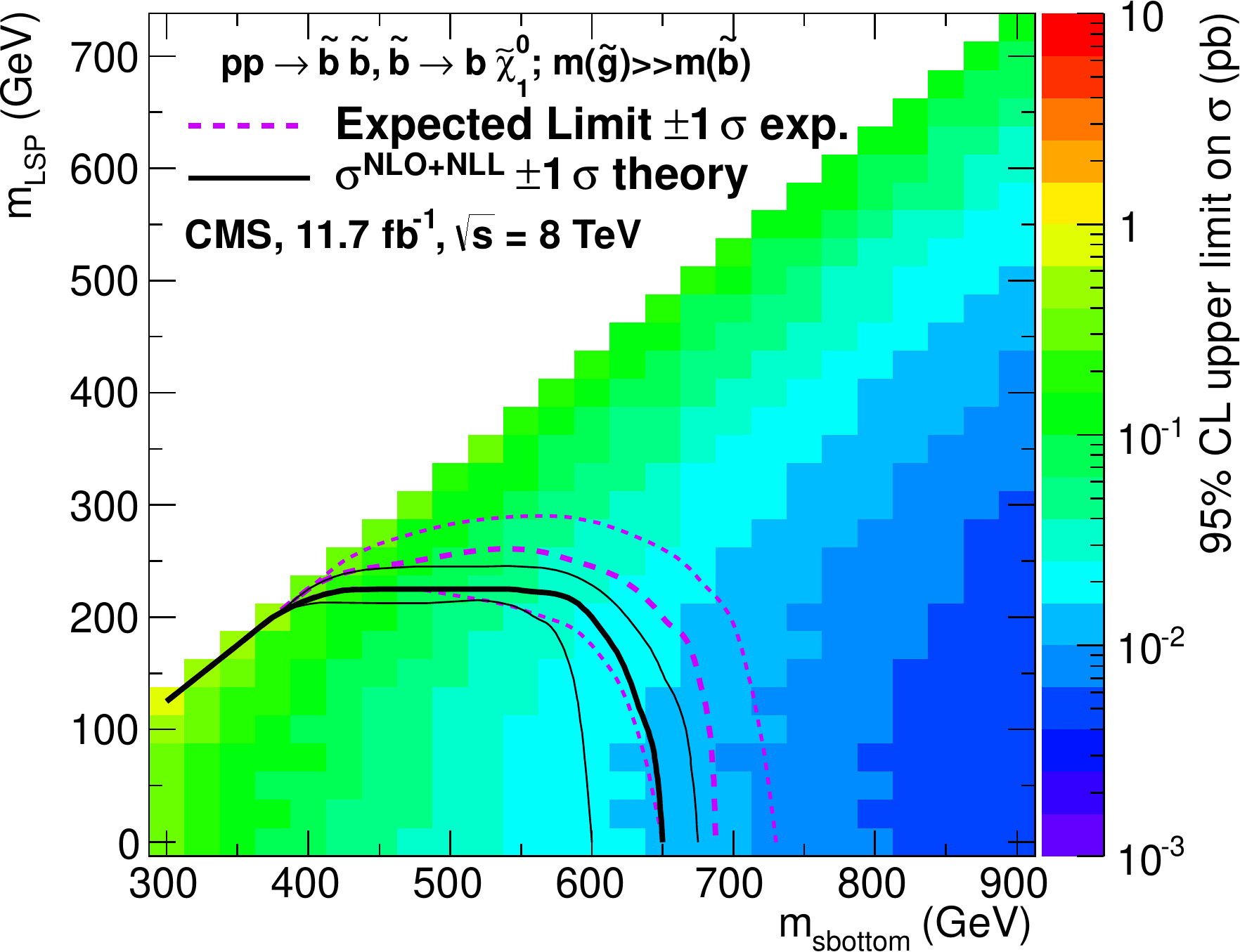} \quad
\includegraphics[width=0.35\linewidth]{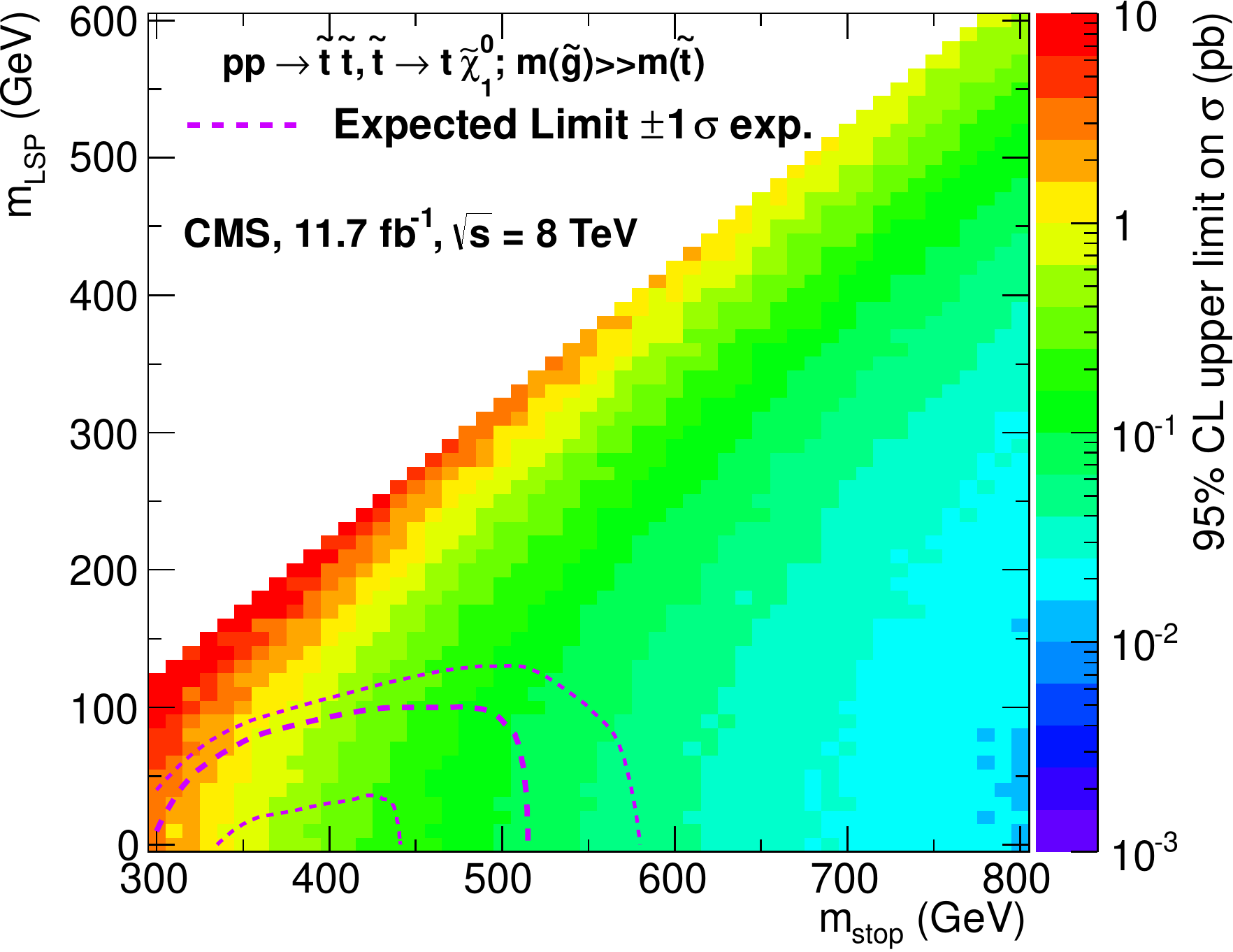}
\caption{95\% CL exclusion upper limit on the cross section times branching ratio versus model spectrum for pair-produced gluinos decaying to two $b$ quarks and a neutralino LSP (top left) or two top quarks and a neutralino LSP (top right), pair-produced bottom squarks decaying to a $b$ quark and a neutralino LSP (bottom left), and pair-produced top squarks decaying to a top quark and a neutralino LSP (bottom right).  The black and red contours show the boundaries where the observed and expected exclusion cross section matches the NLO+NLL cross section computed for the SUSY model, along with the $\pm 1\sigma$ experimental uncertainties.}
\label{fig:RA1interp}
\end{center}
\end{figure}

In the absence of a signal, limits are set using simplified model spectra (SMSs)~\cite{Alves:2011wf}, which consist of a set of particles and sequence of their productions and decays.  95\% confidence level upper limit on the product of cross section times branching ratio is derived as a function of sparticle masses.  Figure~\ref{fig:RA1interp} shows the exclusion limits on the gluino-LSP plane where pair produced gluinos decay fully to two $b$ quarks plus LSP or two top quarks plus an LSP, on the sbottom-LSP plane where pair produced sbottoms decay fully to $b$ plus LSP, and on the stop-LSP plane where pair-produced stops decay fully to top plus LSP.  This search was able to probe gluinos, sbottoms and stops with masses up to 1.2\,TeV, 650\,GeV and 500\,GeV and cross section times branching ratios down to 10\,fb, 20\,fb and 100\,fb respectively.

\section{jets, $b$-jets, $E_T^{miss}$ with variables $H_T$ and $\Delta\phi_{min}$}

The second inclusive search~\cite{Chatrchyan:2013wxa} directly probes SUSY with 3rd generation and investigates hadronic final states with at least one $b$-tagged jet and large \met using 19.4\,fb$^{-1}$ of data.  This search uses the minimum normalized azimuthal angle \dpm between the \met vector and one of the three highest $p_T$ jets defined as $\Delta \phi_{min} \equiv \Delta \phi_i / \sigma_{\Delta \phi, i}$ as a discriminator between SUSY signal with genuine \met and background processes with fake \met.  The normalization factor $\sigma_{\Delta \phi, i}$ is the resolution of jet $i$ defined as $\sigma_{\Delta \phi_i} = \arcsin (\sigma_T/E_T^{miss})$ where $\sigma_{T_i}$, the uncertainty on the component of the \met vector perpendicular to jet $i$ is calculated from the resolutions of all remaining jets $k$, $\alpha_{p_T} = 0.1p_T$, and angles $\alpha_k$ between jets $k$ and the direction opposite jet $i$ as $\sigma_{T_i}^2 \equiv \sum_k (\sigma_{p_T, k} \sin \sigma_k)^2$.  

The principle SM backgrounds in this final state arise from the production of events with a top quark-antiquark ($\bar{t}t$) pair, a single top quark, $W$ or $Z$ bosons in association with jets $(W/Z+{\rm jets})$ and multijets produced through strong interaction in which a $b$-tagged jet is present.  Data were collected using triggers with requirements on thresholds of $H_T$ and missing hadronic transverse momentum $H_T^{miss}$.  Events were required to have at least one good primary vertex, at least 3 jets with $p_T > 50$\,GeV, $|\eta| < 2.4$ where first and second jet $p_T > 70$\,GeV, at least one $b-$tagged jet, $H_T > 400$\,GeV and $E_T^{miss} > 125$\,GeV.  Events having isolated electrons with $p_T > 10$\,GeV, $|\eta| < 2.5$; isolated muons with $p_T > 10$\,GeV, $|\eta| < 2.4$ and isolated charged tracks with $p_T > 15$\,GeV, $|\eta| < 2.4$ were vetoed.  Signal sample is defined by the requirement of $\Delta \hat{\phi}_{min} > 4.0$.  

\begin{figure}[htbp]
\begin{center}
\includegraphics[width=0.32\linewidth]{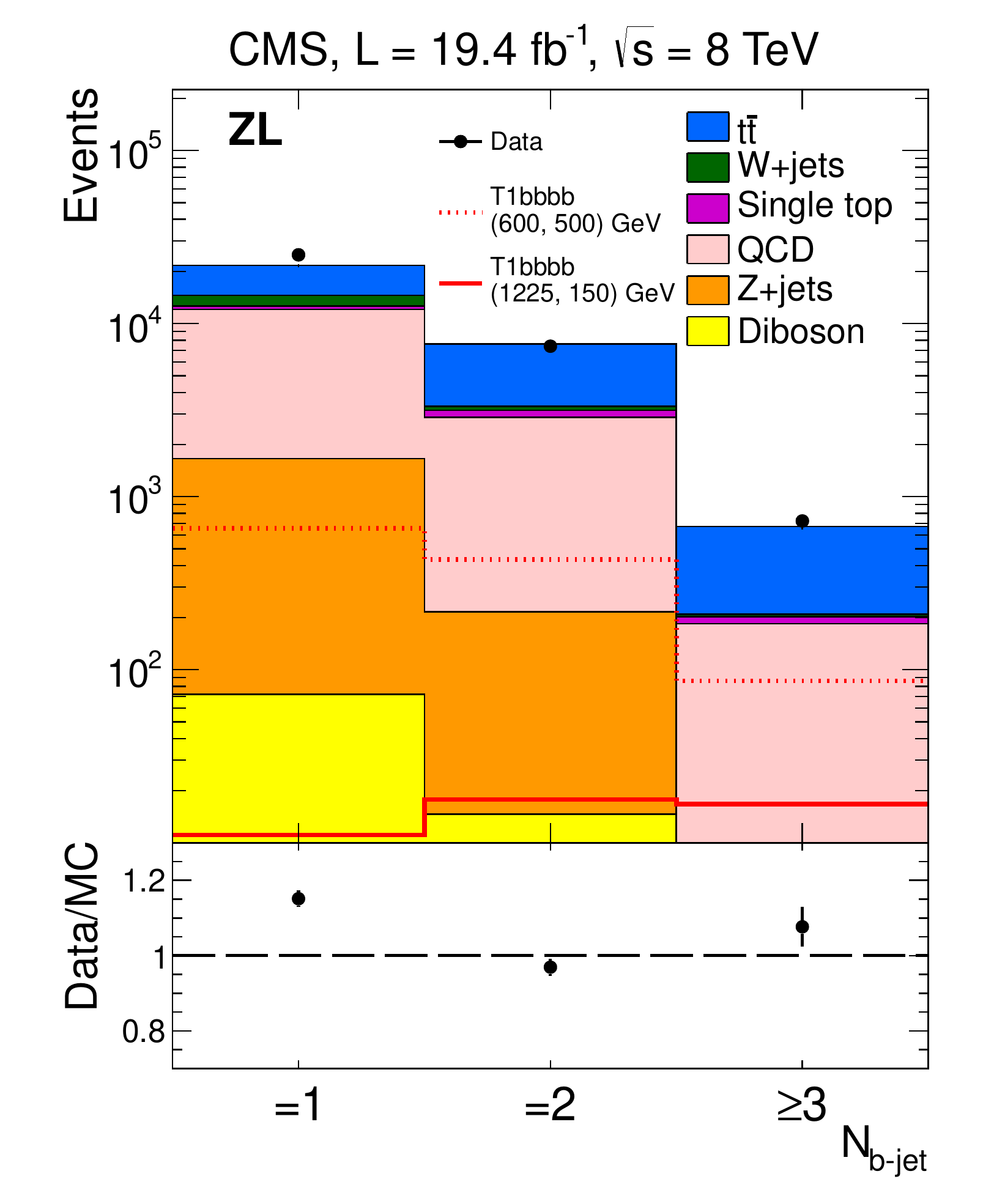}
\includegraphics[width=0.32\linewidth]{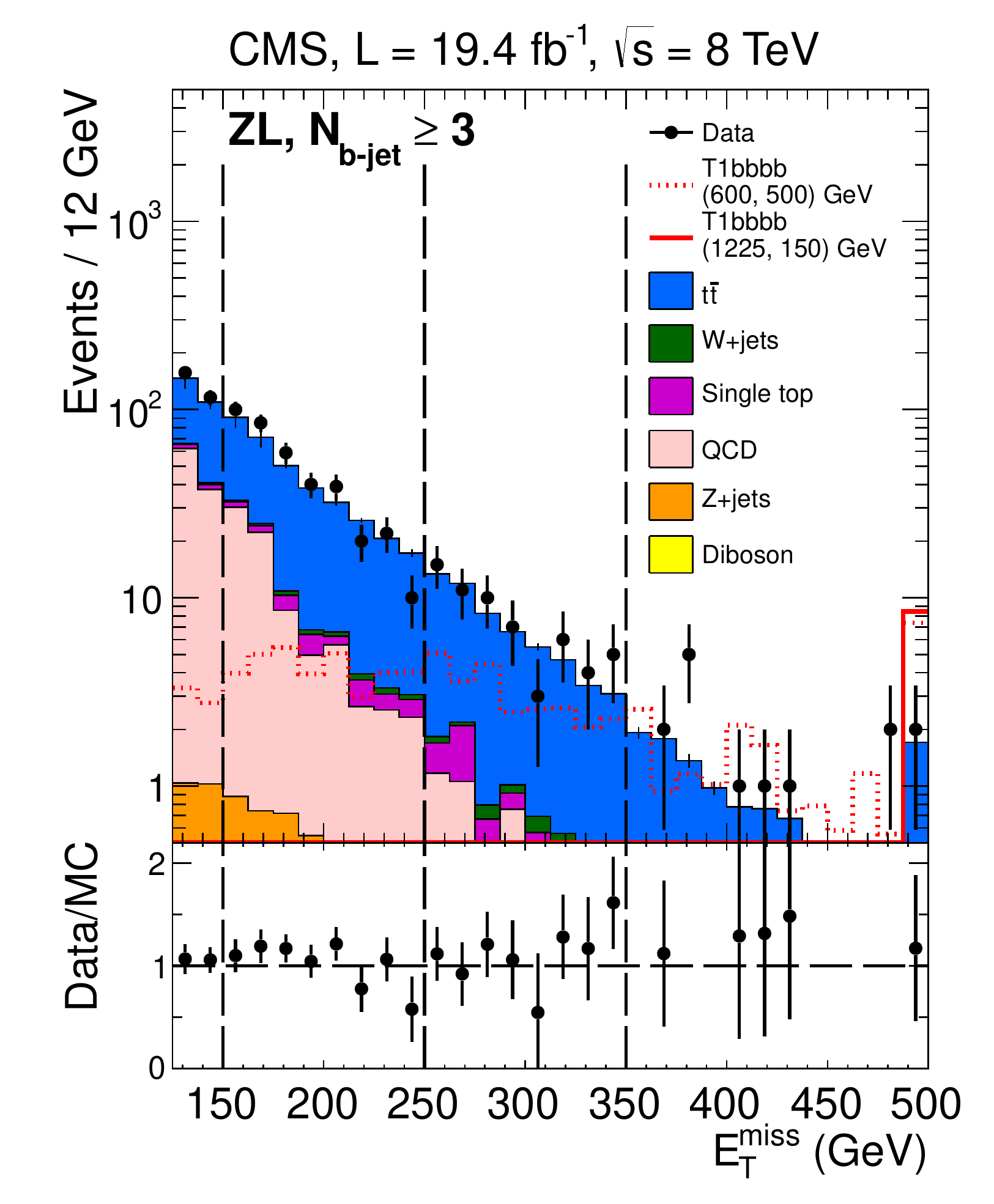}
\includegraphics[width=0.32\linewidth]{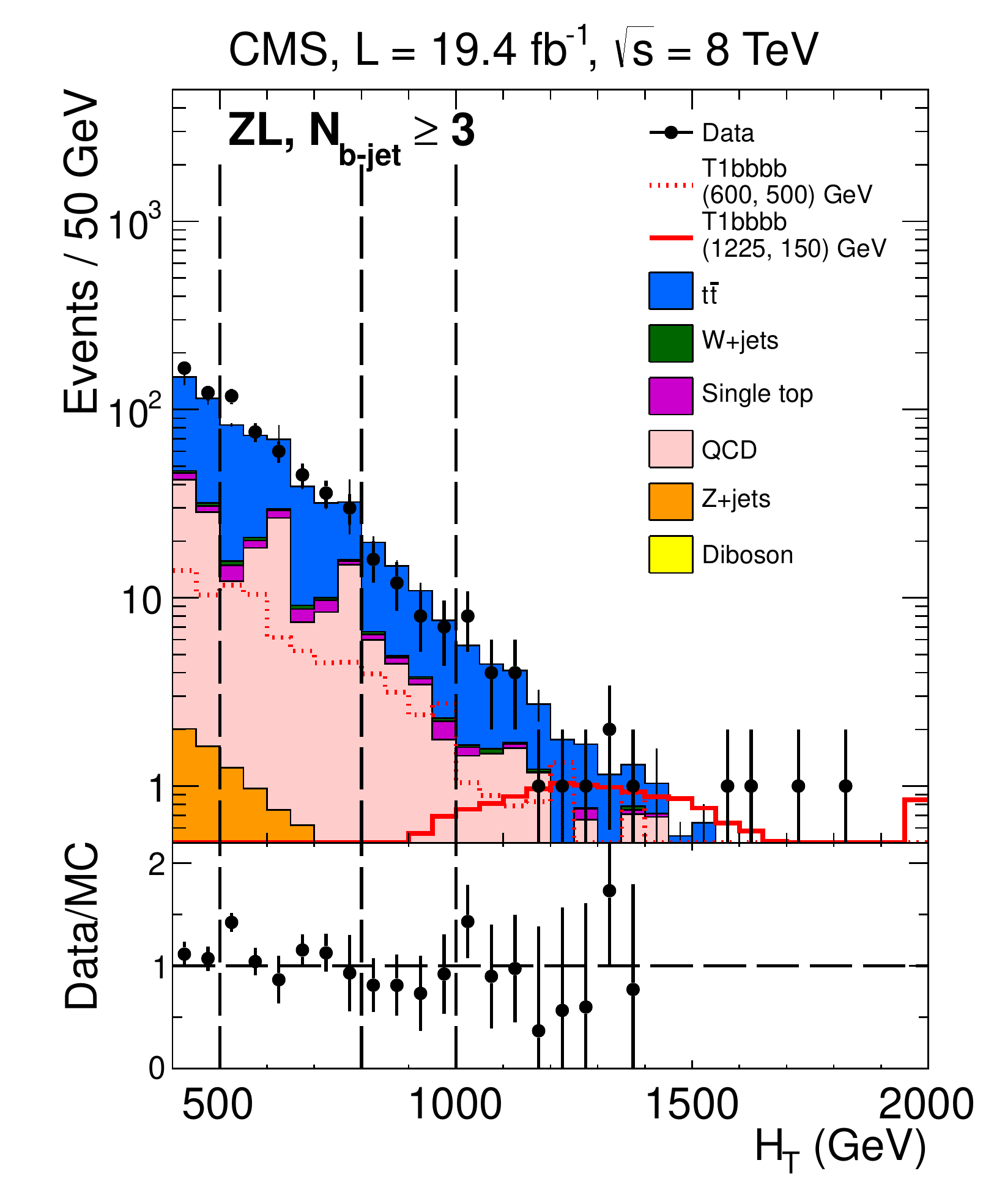}
\caption{Data and Monte Carlo distributions of $N_{b-jet}$ (left), \met (middle) and $H_T$ (right) for the signal sample. The lower panes show the ratio of the measured to the simulated events. The simulated results are intended for guidance and are not used in the analysis. Results for the T1bbbb scenario with (m$\tilde{g}$, m$\tilde{\chi}^0_1$)=(600 GeV, 500 GeV) and (1225 GeV, 150 GeV) are shown as unstacked distributions. The uncertainties are statistical only. The normalization of the simulated curves is based on the absolute cross sections.}
\label{fig:RA2bdatamc}
\end{center}
\end{figure}

The selected events are divided into mutually exclusive bins in $H_T$, \met and $b$-jet multiplicity, to look for a signal excess.  Figure~\ref{fig:RA2bdatamc} shows the distributions in the signal region for $b$-jet multiplicity $N_{b-jet}$, \met and $H_T$ for data, SM MC and two signal benchmarks, where $H_T$ and \met distributions are shown for $N_{b-jets} = 3$.  SM MC distributions agree well with observed data, however a mostly data-driven method similar to that employed by the $\alpha_T$ search was used for a more reliable estimation of the background contamination in the signal sample.  A control sample for top and $W+{\rm jets}$ backgrounds was obtained by reverting the isolated lepton veto in the signal selection and requiring leptons, a control sample for QCD multijet events was obtained by reverting the \dpm cut, and a control sample for $Z(\rightarrow \nu \nu)+{\rm jets}$ was obtained from $Z(\rightarrow ee/\mu\mu)+{\rm jets}$ events. All control samples were binned like the signal sample and expected background events in each bin of the signal sample was estimated by multiplying the observed yield in the control sample relevant for that background with bin-by-bin scale factors estimated from the MC.

\begin{figure}[htbp]
\begin{center}
\includegraphics[width=0.50\linewidth]{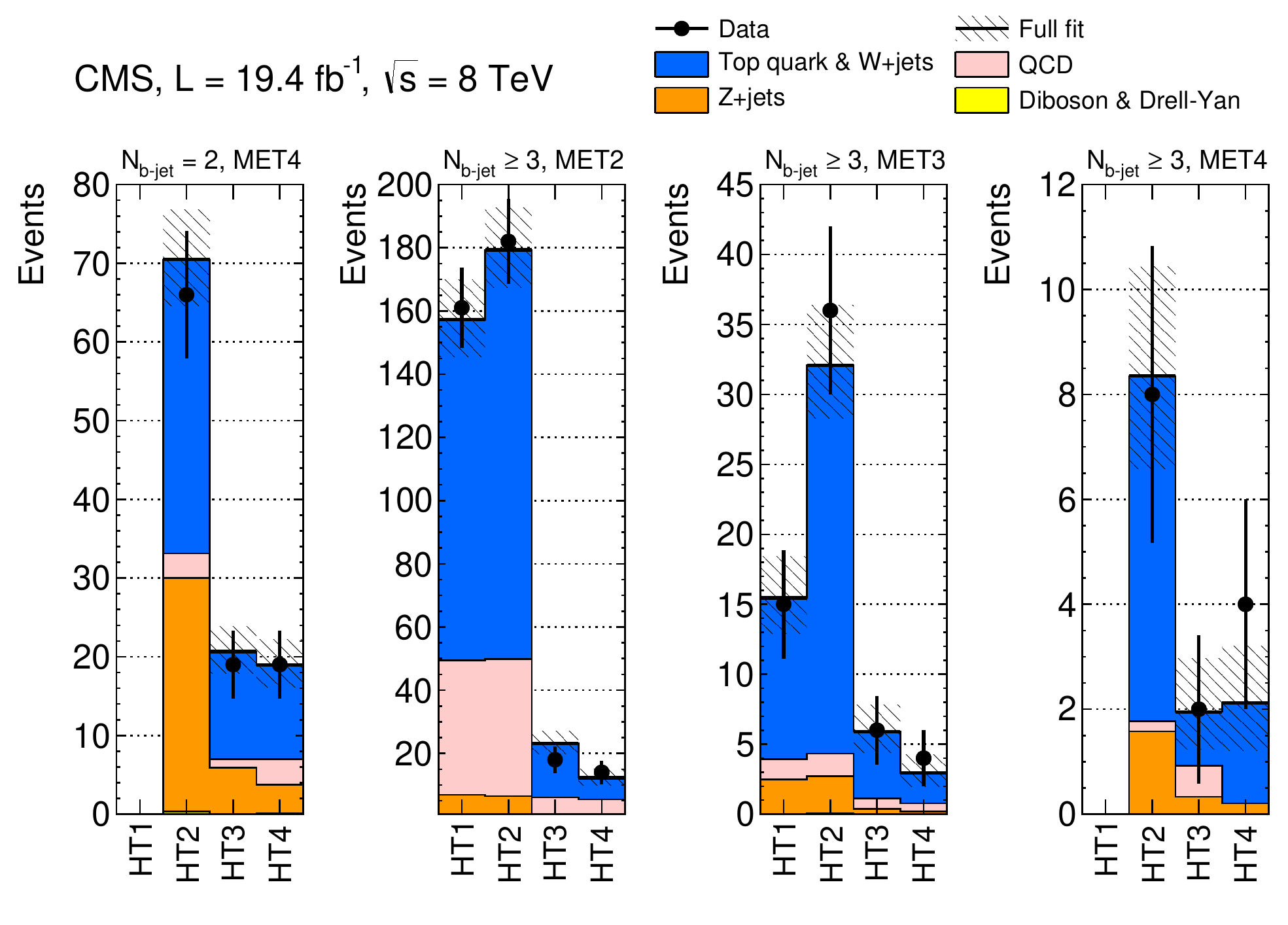}
\caption{Observed numbers of events for the 14 bins with highest signal sensitivity in the analysis, in comparison with the SM background predictions (with total uncertainties) found in the fit with SUSY signal strength fixed to zero. The labels HT1, HT2, MET2, etc., refer to the bins of $H_T$ and \met .}
\label{fig:RA2bfitw0sig}
\end{center}
\end{figure}

The expressions relating signal and control samples, the observed quantities and the parameters were all put together in a likelihood, which is a product of Poisson probabilities, one for each bin, and the constraint PDFs for the nuissance parameters, and the likelihood was minimized to obtain parameter values.  Data were found to be consistent with the SM.  Figure~\ref{fig:RA2bfitw0sig} shows the observed number of events in the 14 bins with the highest signal sensitivity in comparison with the SM background predictions found by performing a fit by setting SUSY signal strengths to zero.  We see that the zero signal expectation agrees very well with the data.

\begin{figure}[htbp]
\begin{center}
\includegraphics[width=0.35\linewidth]{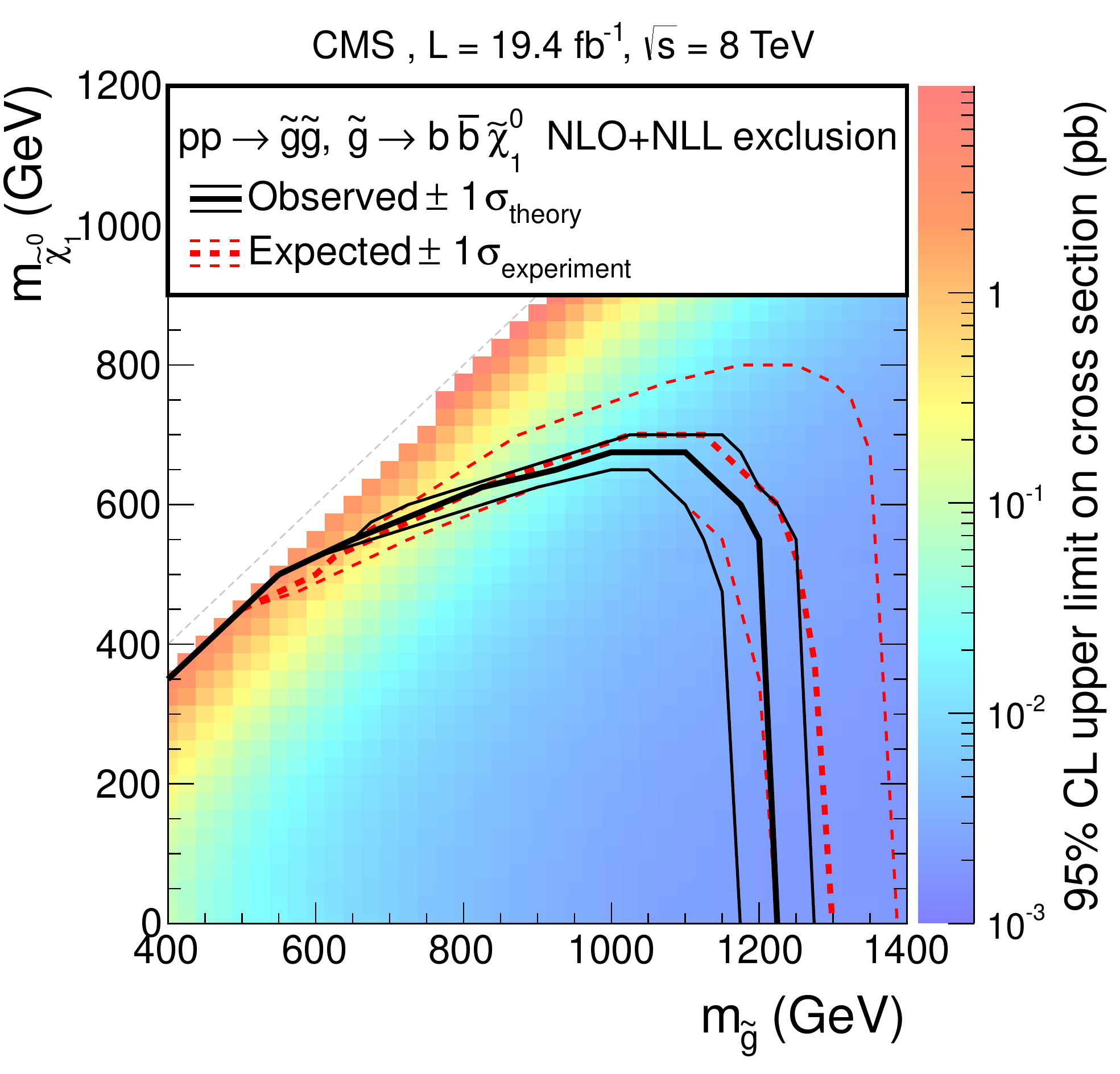} \quad
\includegraphics[width=0.35\linewidth]{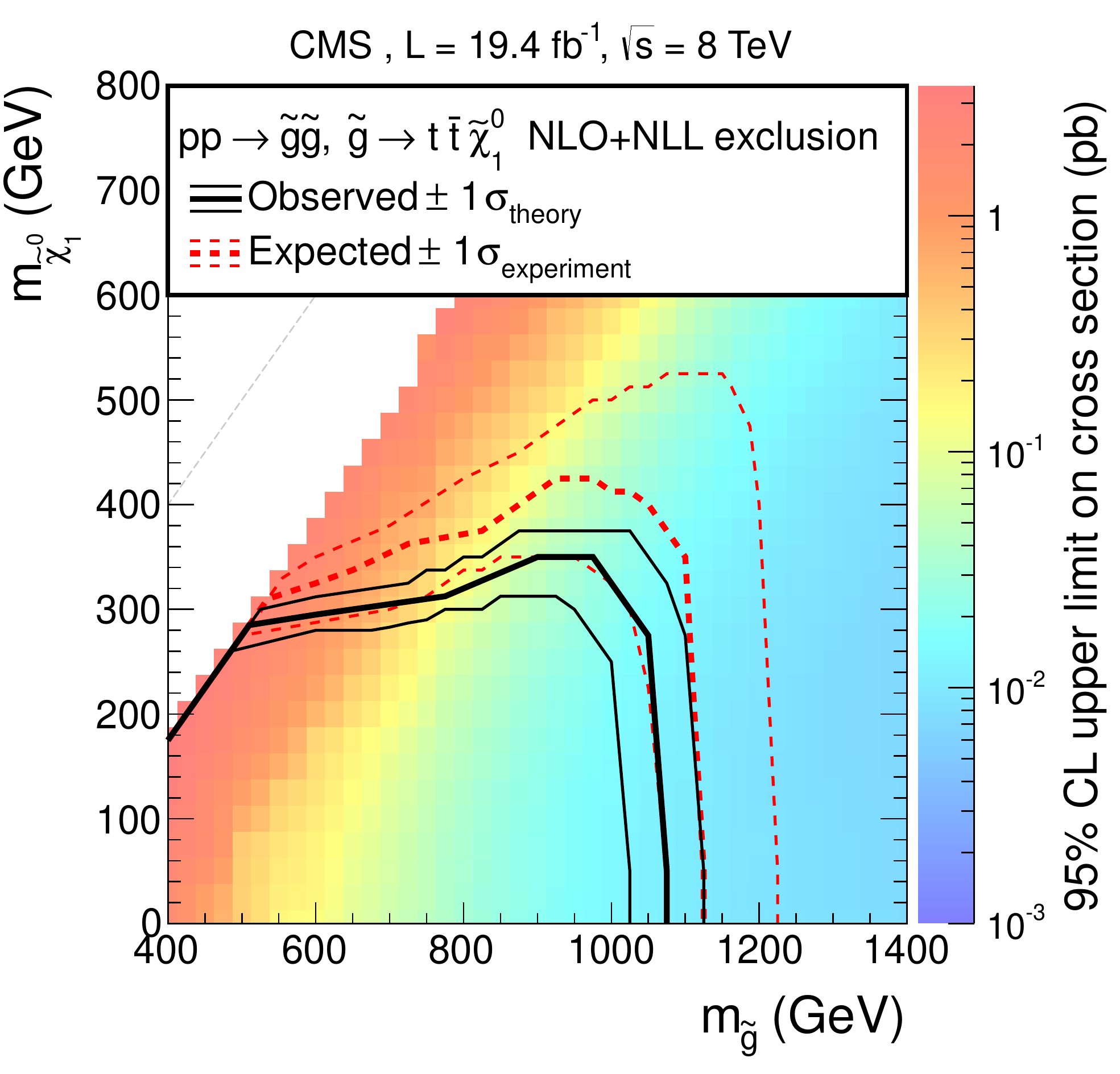} \\
\includegraphics[width=0.35\linewidth]{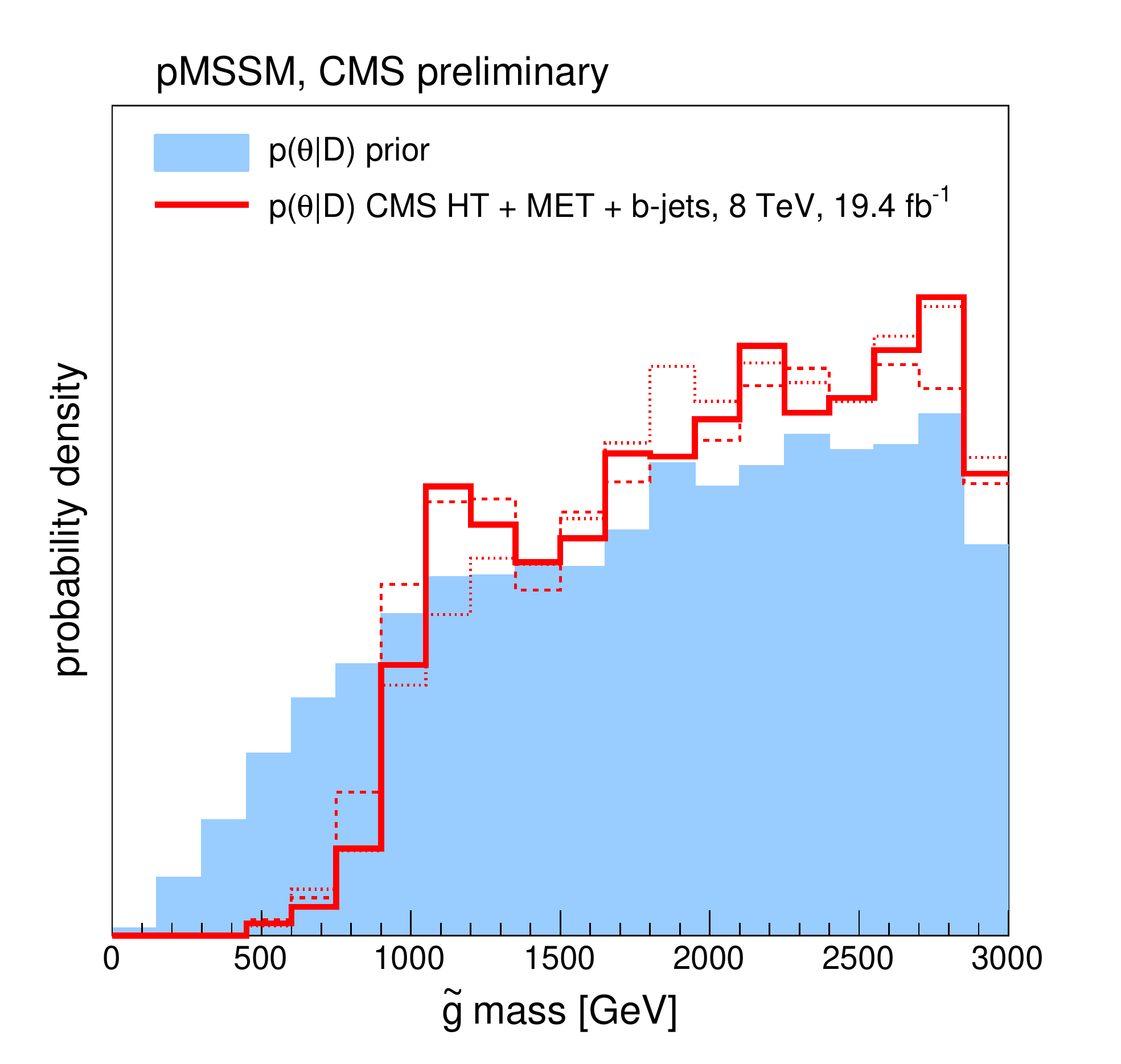} \quad
\includegraphics[width=0.35\linewidth]{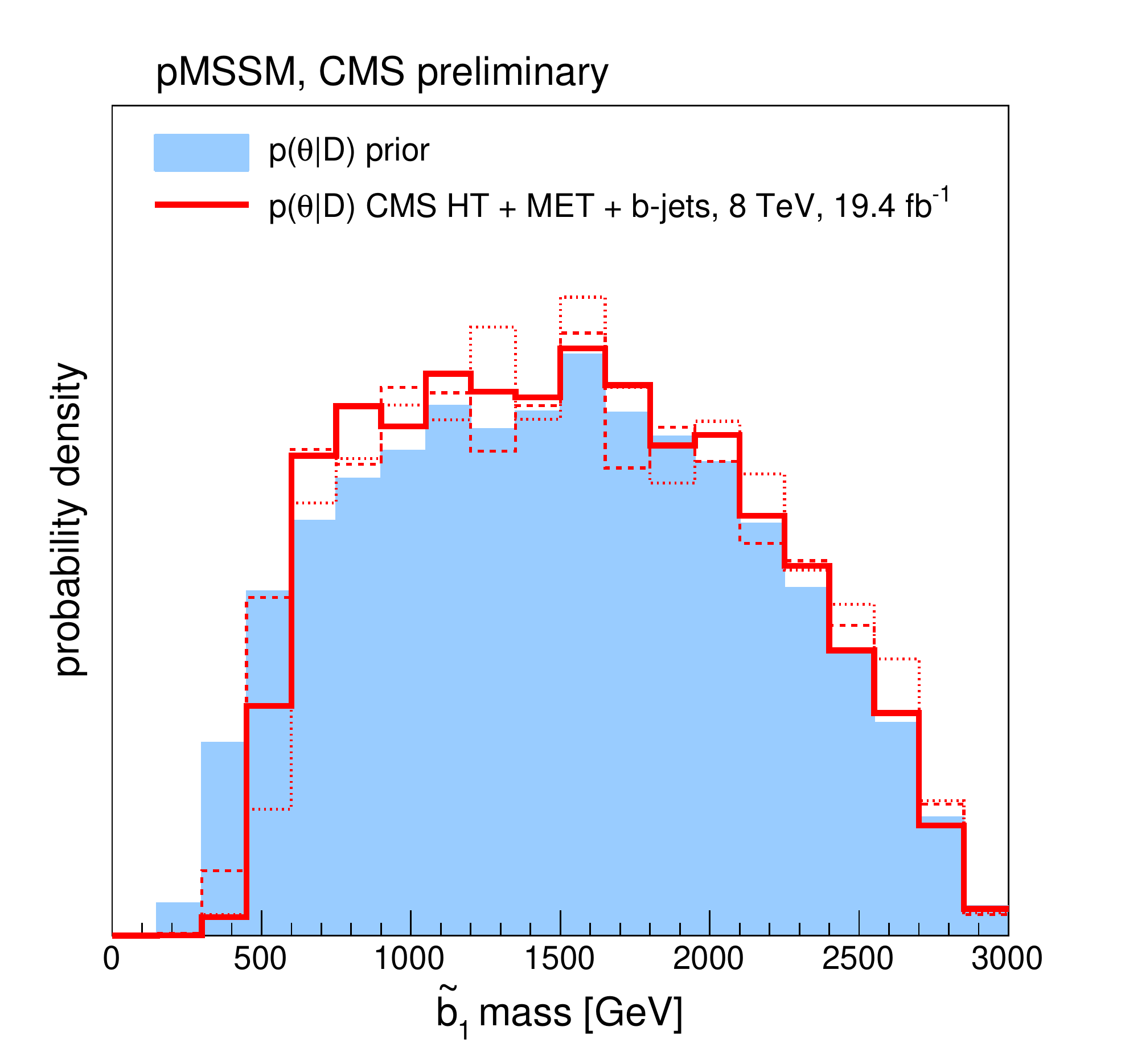}
\caption{95\% CL exclusion upper limit on the cross section times branching ratio versus model spectrum for pair-produced gluinos decaying to two $b$ quarks and a neutralino LSP (top left) and two top quarks and a neutralino LSP (top right).  The black and red contours show the boundaries where the observed and expected exclusion cross sections match the NLO+NLL cross section computed for the SUSY model, along with the $\pm 1\sigma$ experimental uncertainties. Marginalized 1D posterior probability distributions for gluino mass (bottom left) and lighter sbottom mass (bottom right) in pMSSM.  Filled blue histograms show the distributions after some preCMS constraints (described in~\cite{SUS12030}), and red histograms show the distributions after including the $H_T$, \met , $b$-jets, \dpm analysis, where the solid, dashed and dotted lines show likelihoods calculated using the central values of estimated signal counts $s$ and values with uncertainties $s-0.5s$ and $s+0.5s$ respectively.}
\label{fig:RA2binterp}
\end{center}
\end{figure}

The impact of this absence was evaluated on simplified models with gluino pair production with gluinos decaying to various final states, and on phenomenological MSSM (pMSSM), a 19-dimensional parameterization of MSSM at the SUSY scale.  Figure~\ref{fig:RA2binterp} shows the 95\% exclusion upper limit on the cross section times BR on the gluino-LSP mass plane among with the exclusion boundaries as well as the impact of this analysis on gluino and light sbottom mass probability distributions for pMSSM.  In SMSs, which are defined as pure initial and final states, gluinos going to $b$s and LSP can be probed up to masses 1.2\,TeV and down to cross section times BRs of ${\cal O}(10^{-3})$\,fb$^{-1}$.  In a generic model like pMSSM, this search is able to make an impact on gluino and sbottom mass distributions, and exclude the full phase space with gluinos up to mass 500\,GeV and sbottoms up to mass 300\,GeV.

\section{Conclusion}

CMS has been looking for natural SUSY with inclusive searches in final states with multiple jets, high missing transverse energy and $b$-tagged jets, using kinematic variables such as hadronic transverse momentum $H_T$, $\alpha_T$ and $\Delta \hat{\phi}_{min}$.  With up to 19.5\,fb$^{-1}$ of 8\,TeV pp data, no excess over the Standard Model has been observed.  Impact of this absence has been studied on simplified models and phenomenological MSSM, and it was shown that current searches have probed gluinos with masses up to 1.2\,TeV, sbottoms with masses up to 650\,GeV and stops with masses up to 500\,GeV.

%\section*{Acknowledgements}

%I thank my colleagues who performed the analyses described in this note.

\nobreak

\end{document}